\documentclass[iop]{emulateapj}

\usepackage{hyperref}
\usepackage{xcolor}
\usepackage{url}
\usepackage{amssymb}
\hypersetup{
    colorlinks,
    linkcolor={red!50!black},
    citecolor={blue!50!black},
    urlcolor={black} 
}

\newcommand{\feh}{\ensuremath{[\mbox{Fe}/\mbox{H}]}}

\shorttitle{Thermal Spectrum of $\mu$ And b}
\shortauthors{Piskorz et al.}

\begin{document}

\title{Detection of Water Vapor in the Thermal Spectrum of \\ the Non-Transiting Hot Jupiter upsilon Andromedae b}

\author{Danielle Piskorz\altaffilmark{1}, Bj{\"o}rn Benneke\altaffilmark{1,2}, Nathan R. Crockett\altaffilmark{1}, Alexandra C. Lockwood\altaffilmark{3}, \\Geoffrey A. Blake\altaffilmark{1}, Travis S. Barman\altaffilmark{4}, Chad F. Bender\altaffilmark{5}, John S. Carr\altaffilmark{6}, John A. Johnson\altaffilmark{7}}

\altaffiltext{1}{Division of Geological and Planetary Sciences, California Institute of Technology, Pasadena, CA 91125}
\altaffiltext{2}{Institute for Research on Exoplanets,  Universit{\'e} de Montr{\'e}al, Montreal, Canada}
\altaffiltext{3}{Science and Technology Corporation, Columbia, MD 21046}
\altaffiltext{4}{Lunar and Planetary Laboratory, University of Arizona, Tucson, AZ 85721}
\altaffiltext{5}{Department of Astronomy and Steward Observatory, University of Arizona, Tucson, AZ 85721}
\altaffiltext{6}{Naval Research Laboratory, Washington, DC 20375}
\altaffiltext{7}{Harvard-Smithsonian Center for Astrophysics; Institute for Theory and Computation, Cambridge, MA 02138}

\begin{abstract}
The upsilon Andromedae system was the first multi-planet system discovered orbiting a main sequence star. We describe the detection of water vapor in the atmosphere of the innermost non-transiting gas giant ups~And~b by treating the star-planet system as a spectroscopic binary with high-resolution, ground-based spectroscopy.  We resolve the signal of the planet's motion and break the mass-inclination degeneracy for this non-transiting planet via deep combined flux observations of the star and the planet. In total, seven epochs of Keck NIRSPEC $L$ band observations, three epochs of Keck NIRSPEC short wavelength $K$ band observations, and three epochs of Keck NIRSPEC long wavelength $K$ band observations of the ups~And~system were obtained. We perform a multi-epoch cross correlation of the full data set with an atmospheric model. We measure the radial projection of the Keplerian velocity ($K_P$ = 55 $\pm$ 9 km/s), true mass ($M_b$ = 1.7 $^{+0.33}_{-0.24}$ $M_J$), and orbital inclination \big($i_b$ = 24 $\pm$ 4$^{\circ}$\big), and determine that the planet's opacity structure is dominated by water vapor at the probed wavelengths. Dynamical simulations of the planets in the ups~And~system with these orbital elements for ups~And~b show that stable, long-term (100 Myr) orbital configurations exist. These measurements will inform future studies of the stability and evolution of the ups~And~system, as well as the atmospheric structure and composition of the hot Jupiter.
\end{abstract}


\section{Introduction}
The first exoplanet in the upsilon Andromedae system was discovered in 1997 with the radial velocity (RV) technique \citep{Butler1997}. Two more years of RV observations revealed the presence of two additional planets in the system, making ups~And~the first multiple exoplanet system discovered around a main sequence star \citep{Butler1999}. Three planets orbit the F star ups~And~A: (1) ups~And~b, a hot Jupiter with a minimum mass of 0.71M$_J$ and a period of  4.617 $\pm$ 0.0003 days, (2) ups~And~c, a gas giant with a minimum mass of 2.11M$_J$ orbiting with a period of 241.2 $\pm$ 1.1 days and an eccentricity of 0.18 $\pm$ 0.11, and (3) ups~And~d, another gas giant having a minimum mass of 4.61M$_J$ orbiting with a period of 1266.6 $\pm$ 30 days and an eccentricity of 0.41 $\pm$ 0.11. Adding to the intrigue, in 2002, a red dwarf companion ups~And~B with a projected separation of 750 AU from ups~And~A was detected and determined to have negligible effects on RV observations \citep{Lowrance2002}. 

This unique assemblage spurred a torrent of investigations into the origin and stability of the system, a few of which we mention here. \cite{Adams2006} showed that the inclusion of general relativity was required to explain the short period and small eccentricity of ups~And~b. Were it not for general relativity, ups~And~b would precess slowly and its eccentricity would be pumped by the massive outer planets. Depending on the mutual inclinations of the planets in the system, it is possible that the Kozai-Lidov mechanism is responsible for the short-period orbit of ups~And~b \citep{Nagasawa2008}, while \cite{Lissaeur2001} suggested that the present-day dynamics of ups~And~b may be detached from that of the outer planets. \cite{Chiang2002} suggested that if the orbital planes of ups~And~c and d were coplanar and locked in an apisidal resonance, then the eccentricity of ups~And~d would be pumped over time as the apsidal resonance damped. Once the apsides are aligned, secular interactions would cause eccentricity to be transferred from ups~And~d to ups~And~c. \cite{Barnes2006} determined that ups~And~c and d lie near the separatrix between libration and circulation, though this behavior could not be explained by planet-planet scattering \citep{Barnes2007}. 

For lack of complete ephemerides, many of these works assumed the planets' minimum masses were their true masses in their models, and therefore that the system was coplanar. A notable exception was \cite{Rivera2000} who concluded that scattering or ejections is a likely cause of the outer planets' high eccentricities. In all, one statement can summarize many of these works: the ups~And~A system is on the edge of instability.

Determining the masses and inclinations of ups~And~A's planets is critical for realistic interpretations of the system's origin and stability. Five 24 $\mu$m Spitzer observations of ups~And~b suggested $i_b>$ 30$^{\circ}$ \citep{Harrington2006}. To that, \cite{Crossfield2010} added seven individual and twenty-eight continuous hours of 24 $\mu$m Spitzer obervations to further constrain $i_b>$ 28$^{\circ}$. This work also reported that the flux maximum for ups~And~b occurred 80$^{\circ}$ before opposition, an observation inconsistent with atmospheric circulation models.

\cite{McArthur2010} used a combination of high-precision astrometry taken with the Fine Guidance Sensor on the Hubble Space Telescope and a large RV data set (974 observations taken over fourteen years) to determine all the orbital elements of ups~And~c and d and provide some insight into the orbital elements of  ups~And~b. ups~And~c was shown to have a mass of 14M$_J$ and inclination of 8$^{\circ}$ from face-on while ups~And~d has a mass of 10$M_J$ and an inclination of 24$^{\circ}$ from face-on. (See Table~\ref{systemproperties} for all reported orbital elements with error bars.) The mutual inclination of  ups~And~c and d is about 30$^{\circ}$. \cite{McArthur2010} made no astrometric detection of ups~And~b, indicating that its inclination must be greater that 1.2$^{\circ}$. They also postulated the presence of a fourth planet in the system in resonance with the third planet and determined that the stellar companion ups~And~B was indeed bound with a true separation of $\sim$ 9900 AU. The existence of the fourth planet ups~And~d was further supported by \cite{Curiel2011}. A non-Newtonian simulation of the system suggested that ups~And~b had an inclination less than $\sim$60$^\circ$ or greater than $\sim$135$^{\circ}$. 

\begin{deluxetable}{llc}[t]
\tablewidth{0pt}
\tabletypesize{\scriptsize}
\tablecaption{$\mu$ And System Properties}
\tablehead{Property & Value & Ref.} 
\startdata
\sidehead{\textbf{$\mu$ And A}}
Mass, $M_{\star}$ & 1.31 $\pm$ 0.02 $M_{\sun}$ & (1)  \\
Radius, $R_{\star}$ & 1.64 $^{+0.04}_{-0.05}R_{\sun}$ & (1) \\
Effective temperature, $T_{\mathrm{eff}}$ & 6213 $\pm$ 44 K & (2) \\
Metallicity, \feh &0.13 $\pm$ 0.07 & (3) \\
Surface gravity, $\log g$ & 4.25 $\pm$0.06 & (2) \\
Rotational velocity, $v \sin i$ & 9.62 $\pm$ 0.5 km/s & (2) \\
Systemic velocity, $v_{sys}$ & -28.59 km/s & (4) \\
\textit{K} band magnitude, $K_{mag}$ & 2.86 $\pm$ 0.08 & (5) \\

\sidehead{\textbf{$\mu$ And b}}
Velocity semi-amplitude, $K$ & 70.51 $\pm$ 0.37 m/s & (6) \\
Line-of-sight orbital velocity, $K_P$ &55 $\pm$ 9 km/s & (7) \\
Indicative mass, $M\sin(i)$ &  0.69 $\pm$ 0.02$M_J$ & (6) \\
Mass, $M_p$ & 1.7 $^{+0.33}_{-0.24}$ $M_J$ & (7) \\
Inclination, $i$ & 24 $\pm$ 4$^{\circ}$& (7) \\
Semi-major axis, $a$ & 0.0594 $\pm$ 0.0003 AU & (6) \\
Period, $P$ &4.617111 $\pm$ 0.000014 d & (6) \\
Eccentricity, $e$ & 0.012 $\pm$ 0.005 & (6) \\
Argument of periastron, $\omega$ &44.11 $\pm$ 25.56$^{\circ}$ & (6) \\
Time of periastron, $t_{peri}$ &   2450034.05 $\pm$ 0.33 JD &(6) \\
Phase uncertainty, $\sigma_{f+\omega}$ & 0.9$^{\circ}$ & (7) \\

\sidehead{\textbf{$\mu$ And c}}
Mass, $M_p$ & 13.98 $^{+2.3}_{-5.3}M_J$ & (6) \\
Inclination, $i$ & 7.868 $\pm$ 1.003$^{\circ}$ &  (6)  \\
Semi-major axis, $a$ &0.8259 $\pm$ 0.043 AU &  (6) \\
Period, $P$ & 240.9402 $\pm$ 0.047 d & (6)  \\
Eccentricity, $e$ & 0.245 $\pm$ 0.006 & (6)  \\
Argument of periastron$^{\tablenotemark{a}}$, $\omega$ & 10.81 $\pm$ 7.73$^{\circ}$ &  (6)  \\
Longitude of periastron,$\varpi$ & 247.66 $\pm$ 1.76$^{\circ}$ & (6) \\
Longitude of ascending node, $\Omega$ & 236.85 $\pm$ 7.53$^{\circ}$  & (6) \\
Time of periastron, $t_{peri}$ &  2449922.53 $\pm$ 1.17 JD & (6) \\

\sidehead{\textbf{$\mu$ And d}}
Mass, $M_p$ & 10.25 $^{+0.7}_{-3.3}M_J$ &  (6)  \\
Inclination, $i$ & 23.758 $\pm$ 1.316$^{\circ}$ &  (6)  \\
Semi-major axis, $a$ & 2.53 $\pm$ 0.014 AU &  (6) \\
Period, $P$ & 1281 $\pm$ 1.055 d &  (6)  \\
Eccentricity, $e$ & 0.316 $\pm$ 0.006 &  (6)  \\
Argument of periastron$^{\tablenotemark{a}}$, $\omega$ &  248.92 $\pm$ 3.55$^{\circ}$  & (6)  \\
Longitude of periastron, $\varpi$ & 252.99 $\pm$ 1.31$^{\circ}$ & (6) \\
Longitude of ascending node, $\Omega$ & 4.07 $\pm$ 3.30$^{\circ}$  & (6) \\
Time of periastron, $t_{peri}$ &  2450059.38 $\pm$ 3.50 JD &  (6)  \\
\enddata
\label{systemproperties}
\tablerefs{(1) \cite{Takeda2007}, (2) \cite{Valenti2005},\\(3) \cite{Gonzalez2007}, (4) \cite{Nidever2002},\\(5) \cite{vanBelle2009}, (6) \cite{McArthur2010},\\(7) This work} 
\tablenotetext{a}{We calculate argument of periastron from the values of longitude of periastron and longitude of ascending node reported in \cite{McArthur2010}. We calculate the error bars on the longitude of periastron by combining the reported error bars on argument of periastron and longitude of ascending node in quadrature.}
\end{deluxetable}

Drawing on the results of \cite{McArthur2010}, \cite{Dietrick2015} ran post-Newtonian numerical simulations of the system to determine which masses and inclinations of ups~And~b would allow the system as a whole to be stable. The system has a general ``region of stability" when $i_b<$ 40$^{\circ}$. Specifically, \cite{Dietrick2015} investigated four stable, prograde simulations having $i_b<$ 25$^{\circ}$, but precise conclusions on the mass and inclination of the innermost planet have eluded astronomers. 

Ground-based high-resolution spectroscopy techniques have successfully broken the degeneracy between mass and inclination for non-transiting planets and would be ideal for determining the mass and inclination of ups~And~b. These techniques treat the target star and its planet as if they were a spectroscopic binary, teasing out the line-of-sight Keplerian velocity of the planet as it orbits the star \citep{Snellen2010}. In addition to untangling the mass and inclinations of bright planets, this technique also gives information on atmospheric composition \citep{Brogi2012, Brogi2013, Brogi2014, deMooij2012, Rodler2012, Birkby2013, deKok2013,  Lockwood2014, Martins2015, Piskorz2016}, wind speed \citep{ Snellen2014}, and length of day \citep{Schwarz2015, Brogi2016} and has been carried out using CRIRES at VLT, HARPS at ESO-La Silla, and NIRSPEC at Keck. Observers using CRIRES (e.g., \citealt{Snellen2010}) or HARPS (e.g., \citealt{Martins2015}) tend to allow the planet lines to smear across the detector over the course of many hours. Observers using NIRSPEC (e.g., \citealt{Lockwood2014}) take up to two hour long snapshots of the planet's emission spectrum at various phases of the planet's orbit. Since NIRSPEC has a resolution of 25,000-30,000 at the observed wavelengths, planet lines generally do not smear across pixels during a 2-3 hour observation. Owing to NIRSPEC's cross-dispersed echelle format, this method yields many planet lines spread over many orders, producing sufficient signal-to-noise to detect the planet's atmosphere. 

In this paper, we use NIRSPEC observations and the methods presented in \cite{Piskorz2016} to discern the true mass, inclination, and atmospheric composition of the hot Jupiter ups~And~b. An important divergence from the method presented in \cite{Piskorz2016} is the inclusion of $K$ band data taken with two different echelle settings, accessing planetary features across the full $K$ band. In Section \ref{methods}, we detail our NIRSPEC observations, data reduction, and telluric correction, while Section \ref{models} describes the cross-correlation analysis and maximum likelihood calculation of the orbital solution for ups~And~b. In Section \ref{discussion}, we discuss the robustness of our orbital solution, the long-term stability of the ups~And~A system, insights into the atmosphere of ups~And~b, and give some notes on the observations.

\section{Observations and Data Reduction}
\label{methods}
\subsection{Observations}
\label{observations}
We used NIRSPEC (Near InfraRed SPECtrometer; \citealt{McLean1998}) at Keck Observatory to observe ups~And~A and b on seven nights (2011 September 6, 7, and 9, 2013 October 27 and 29 and November 7, and 2014 October 7) in $L$, three nights (2016 September 19, November 12, and December 15) in $K_r$ (the \textbf{right}, long-wavelength half of the dispersed, $K$-band filtered light), and three nights (2014 October 5 and 2016 August 21 and September 19) in $K_l$ (the \textbf{left}, short-wavelength half of the dispersed, $K$-band filtered light). We obtained spectral resolutions of $\sim$25,000 in $L$ and $\sim$30,000 in $K$ using the 0.4''x24'' slit setup and used an ABBA nodding pattern during data acquisition. In $L$ band, the echelle orders typically cover 3.4038-3.4565/3.2467-3.3069/3.1193-3.1698/2.995-3.044~$\mu$m.  The echelle orders in $K_r$ band typically cover 2.38157-2.41566/2.31-2.34284/2.24245-2.27485/2.17894-2.20861/2.11878-2.14639/2.06170-2.08703~$\mu$m, while in $K_l$ band the echelle orders typically cover 2.34238-2.37535/2.27198-2.30374/2.20554-2.23653/2.14362-2.17298/2.08461-2.11312/2.02931-2.05634~ $\mu$m. Altogether, the two $K$ band setups provide near continuous wavelength coverage across the entire $K$ band. Table~\ref{observationtable} gives the details of these thirteen nights of observations.

\begin{deluxetable*}{lccccc}
\tablewidth{0pt}
\tablecaption{NIRSPEC Observations of ups~And~b}
\tablehead{Date & Modified Julian Date$^{\tablenotemark{a}}$& Mean anomaly $M^{\tablenotemark{b}}$ & Barycentric velocity $v_{bary}$ & Integration time & S/N$^{\tablenotemark{c}}_{\textit{L}, \textit{K}}$ \\
 & (- 2,400,000.5 days) &  (2$\pi$ rad)  & (km/s) & (min) & }
\startdata
\sidehead{\textbf{\textit{L} band (3.0 - 3.4 $\mu$m)}} 
2011 September 6  & 55810.639 & 0.25 & 21.07 & 60& 5376\\
2011 September 7  & 55811.637 & 0.46 & 20.82 & 10$^{\tablenotemark{d}}$ & 2661\\
2011 September 9  & 55813.509 & 0.87 & 20.33 & 100 & 8265\\
2013 October 27  & 56592.526 & 0.59 & 1.89 & 140 & 9173\\
2013 October 29  & 56594.512 & 0.02 & 0.99 & 140 & 5937\\
2013 November 7  & 56603.609 & 0.99 & -3.17 & 180 & 8686\\
2014 October 7  & 56937.553 & 0.32 & 10.64 & 50 & 5641\\
\sidehead{\textbf{\textit{K$_r$} band (long wavelength side of 2.0 - 2.4 $\mu$m) } }
2016 September 19  & 57650.361 & 0.73 & 17.14 & 100 & 11517\\
2016 November 12  & 57704.265 & 0.38 & -5.37 & 230 & 12872\\
2016 December 15 & 57737.300 & 0.53& -18.63 & 70 & 7666\\
\sidehead{\textbf{\textit{K$_l$} band (short wavelength side of 2.0 - 2.4 $\mu$m) } }
2014 October 5  & 56935.579 & 0.87 & 11.47 & 70 & 7764\\
2016 August 21  & 57621.589 & 0.45 & 24.15 & 30 & 4369\\
2016 September 19  & 57650.501 & 0.73 & 17.14 & 120 & 10649\\
\enddata
\label{observationtable}
\tablenotetext{a}{Julian date refers to the middle of the observing sequence.}
\tablenotetext{b}{We list only the mean anomalies (and no true anomalies) for our observations, since ups~And~b's orbit is nearly circular.}
\tablenotetext{c}{S/N$_{\textit{L}}$, S/N$_{\textit{K}_r}$, S/N$_{\textit{K}_l}$ are calculated at 3.0, 2.1325, and 2.1515 $\mu$m, respectively. Each S/N calculation is for a single channel (i.e., resolution element) for the whole observation.} 
\tablenotetext{d}{As the total integration time on ups~And~on 2011 September 7 is very short, we do not use principal component analysis to remove the telluric signals (see Section~\ref{reduction}), and we exclude this epoch from the following analysis.} 
\end{deluxetable*}

\begin{figure}[t]
\centering
\noindent\includegraphics[width=20pc]{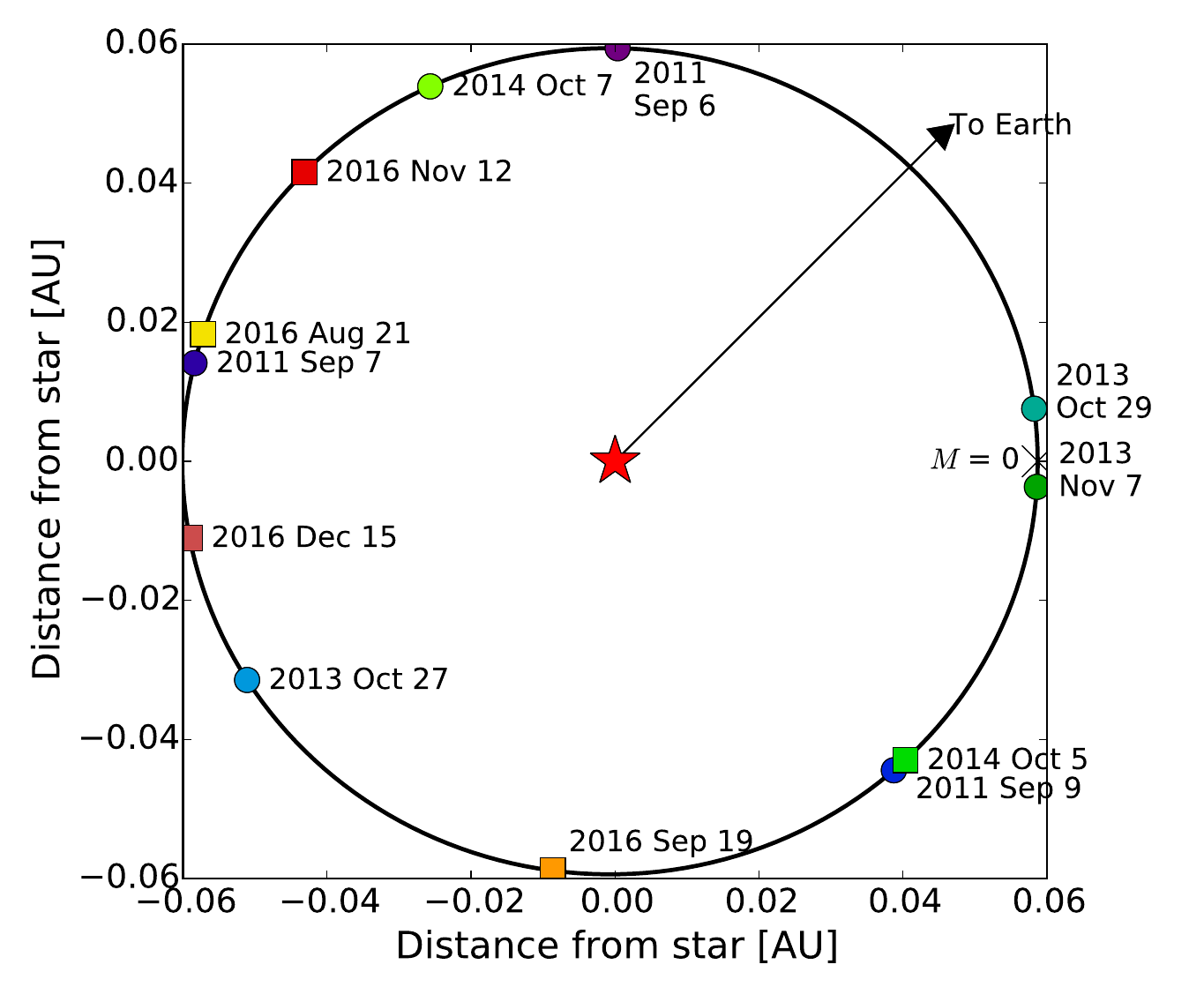}
\caption{Top-down schematic of the orbit of ups~And~b around its star according to the orbital parameters derived by \cite{McArthur2010}. Each point represents a single epoch of NIRSPEC observations of the system. Circles indicate \textit{L} band observations and squares represent \textit{K} band observations. The black arrow represents the line of sight to Earth.}
\label{schematic}
\end{figure}

\begin{figure}[t]
\centering
\noindent\includegraphics[width=20pc]{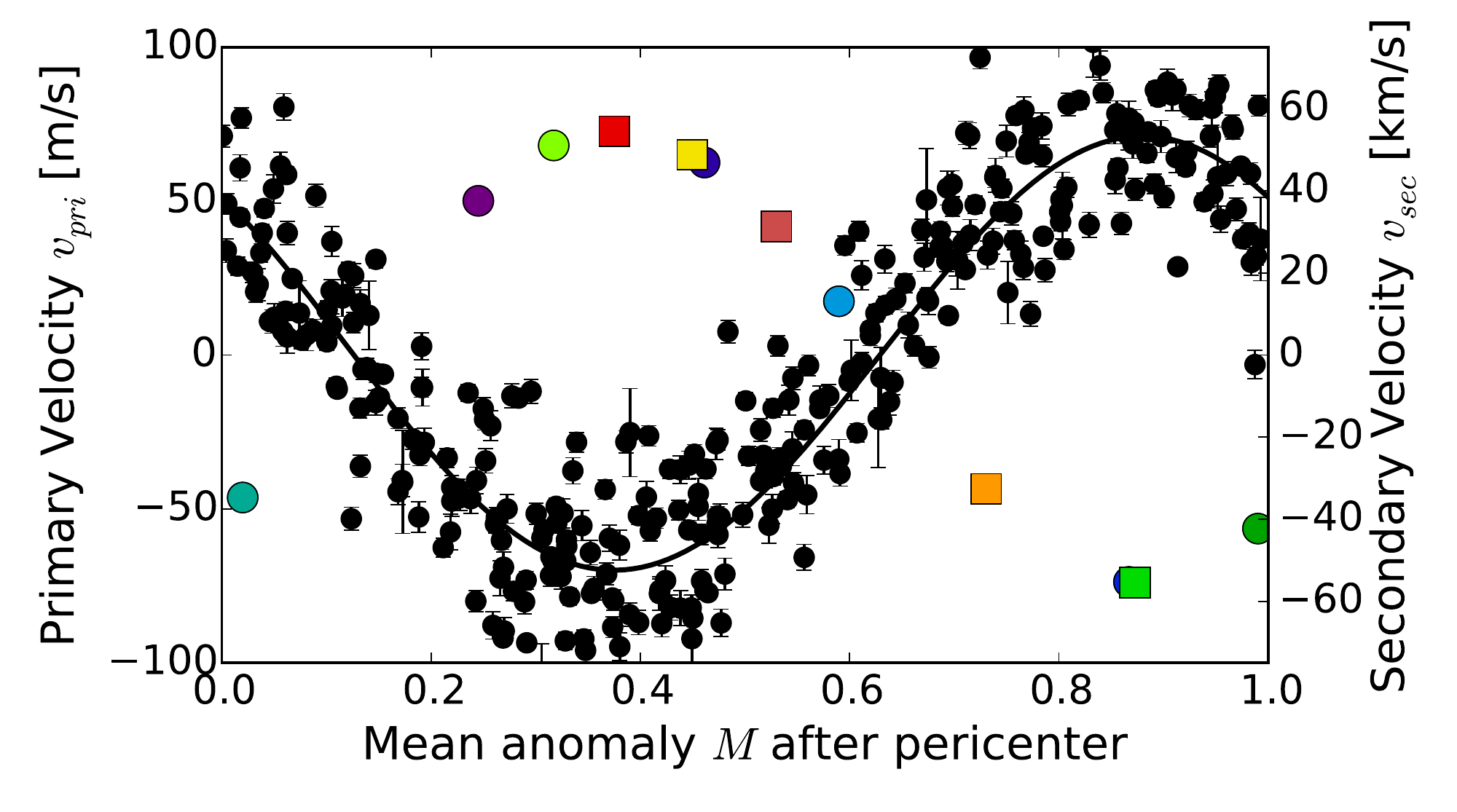}
\caption{RV data from \cite{Fischer2014} with the best-fit stellar RV (primary velocity) curve over-plotted in black, corresponding to the left y-axis. RV contributions from ups~And~c and d have been removed according to the orbital elements provided in \cite{McArthur2010}. The colored points represent the NIRSPEC observations of this planet correspond to the right y-axis and are based on the observation phases and our expectations of their secondary velocities. In the course of this paper, we will show that the most likely value for the Keplerian orbital velocity of ups~And~b is 55 $\pm$ 9 km/s.}
\label{rvplot}
\end{figure}

A top-down schematic of ups~And~b in orbit around ups~And~A is shown in Figure~\ref{schematic} with the expected orbital phase of each observational epoch marked. Figure~\ref{rvplot} shows radial velocity measurements of ups~And~A taken from \cite{Fischer2014} in comparison with expectations for the line-of-sight velocity of ups~And~b. We aim to take observations when the line-of-sight velocities of the star and planet are most distinct and when we expect to observe a decent amount of dayside radiation from the planet, thus maximizing the planet flux. 

\subsection{Extraction of 1-D Spectra and PCA-like Telluric Correction}
\label{reduction}
Our data reduction and cleaning methods are parallel to those described in \cite{Piskorz2016} and are summarized here only briefly.

We use a Python pipeline in the style of \cite{Boogert2002} to flat field and dark subtract our data, remove bad pixels, and extract 1-D spectra. For the wavelength calibration, we fit a fourth-order polynomial ($\lambda = ax^3+bx^2+cx+d$, where $x$ is pixel number and $a, b, c,$ and $d$ are free parameters) that aligns our $L$ band data to a telluric model or our $K$ band data to a combined telluric and stellar model. Here, the difference in treatment of $L$ and $K$ band data stems from the fact that telluric lines are stronger in the $L$ band than near 2 $\mu$m. Our stellar model is derived from the PHOENIX stellar library \citep{Husser2013} and is described in more detail in Section~\ref{stellarmodel}. Finally, we fit an instrument profile to our data as in \cite{Valenti1995} and save it to apply to the models described in Sections~\ref{stellarmodel} and~\ref{planetmodel}.

\begin{figure}[b]
\centering
\noindent\includegraphics[width=21pc]{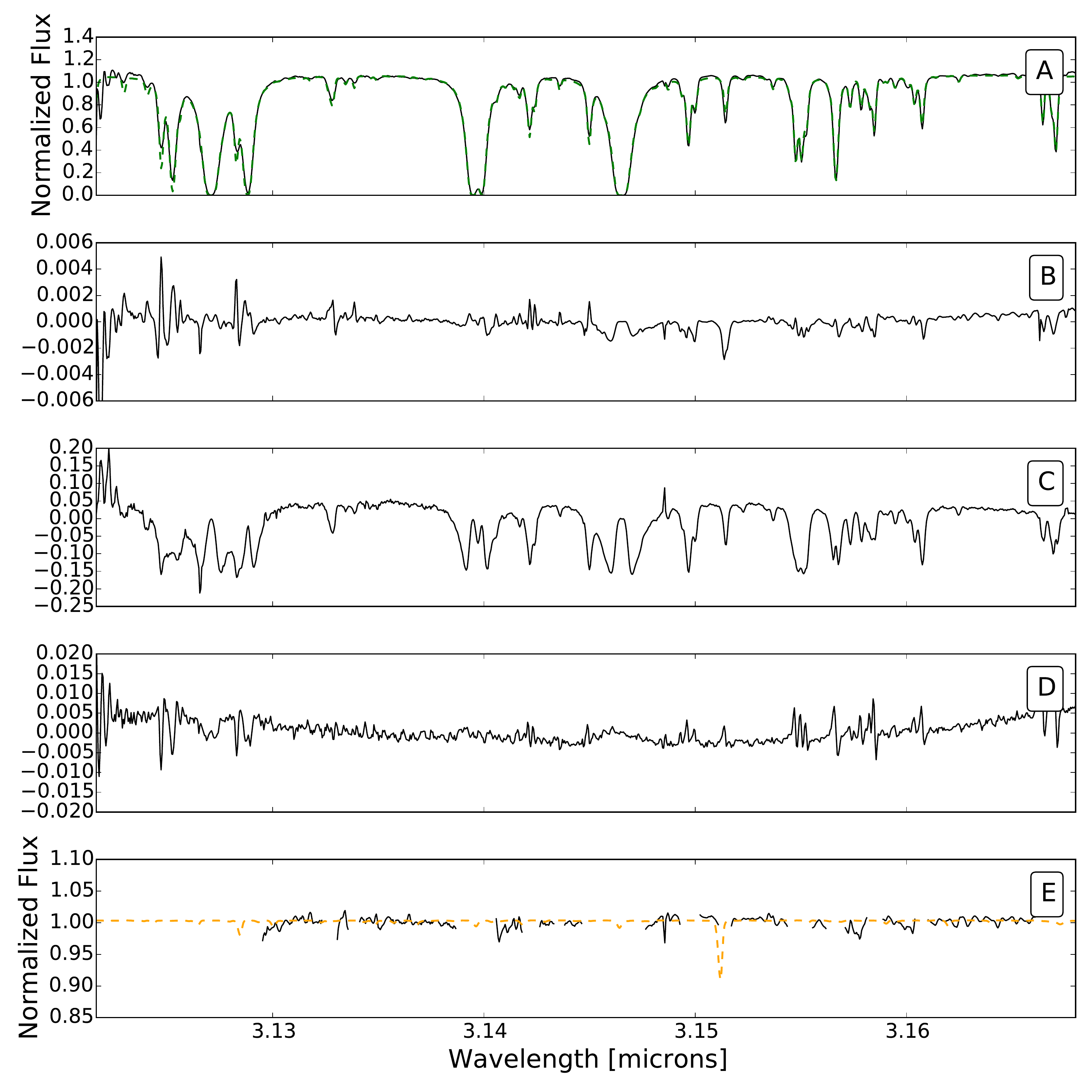}
\caption{Raw spectrum of ups~And, first three principal components, and cleaned spectrum. (A): One order of data from ups~And taken on 2013 October 29. The best-fit telluric spectrum is over plotted as a green, dashed line. (B-D): The first three principal components in arbitrary units describing changes in air mass, molecular abundances in the Earth's atmosphere, and plate scale, respectively. (E): Same as (A), but with the first five principal components removed, and with a fitted stellar spectrum overplotted as a dashed, orange line.}
\label{pcafigure}
\end{figure}

We capitalize on the long time series of observations (roughly two minutes per nod, or four minutes per AB pair) taken at each epoch and perform a principal component analysis (PCA) to remove contributions to the spectra from the Earth's atmosphere. PCA rewrites a data set in terms of its principal components so that the variance of a data set with respect to a model or its mean is reduced. For our purposes, this means that PCA will identify the time-varying components of our time-series data, most notably, changes in the telluric spectrum over the course of a given epoch. The first principal component describes the most variance, the second, the second most, etc. We guide our PCA with a telluric model that best fits the data in terms of water, carbon dioxide, methane, and (where appropriate) ozone abundances, and determine the eigenvectors making up each observed spectrum. We calculate and remove the strongest principal components from our data, leave behind the parts of the spectra which are constant in time (the stellar and planet signals), combine every AB nod of data, and clip regions of substantial telluric absorption ($>$75$\%$). More information on this technique is given in \cite{Piskorz2016}. Figure \ref{pcafigure} shows a raw spectrum of ups~And taken on 2013 October 29, the first three principal components, and a cleaned spectrum of ups~And.

As in our analysis of HD 88133 data, we find the telluric correction by PCA works well for all orders of $L$ band data, but poorly for the $K_r$ and $K_l$ band orders spanning 2.06170-2.08703~$\mu$m and 2.02931-2.05634~ $\mu$m where there is a dense forest of telluric CO$_2$ lines. We also find that a few nights of $K$ band observations were contaminated by significant issues with the read-out electronics. In these cases, we exclude the data on the ``bad" side of the detector from our analysis; about 25\% of the data is on the noisy side of the detector. Additionally, we remove the 2011 September 7 observations from our data set, since the ten minute total integration time is not sufficient for principal component analysis.

As in \cite{Piskorz2016}, all but about 0.1$\%$ of the variance in each night's data set is encapsulated by the first principal component. The following results are roughly consistent for data sets with more than the first principal component removed. As discussed in Section \ref{obsnotes} and shown in Figure \ref{contrast}, the expected photometric contrast $\alpha_{phot}$ at the observed wavelengths is $\sim$10$^{-6}$. Based on the percent variance removed by each principal component we determine that deletion of a signal of this magnitude requires the removal of upwards of the first fifteen principal components from our data. In the analysis that follows, our data set has the first five principal components removed, leaving the stellar and planetary signals intact.

\pagebreak
\section{Data Analysis and Results}
\label{models}
A two-dimensional cross-correlation analysis reveals the ideal velocity shifts for the stellar and planet spectra embedded in our clean data set \citep{Zucker1994}. This analysis calls for accurate stellar and planetary model spectra.

\subsection{Model Stellar Spectrum}
\label{stellarmodel}
Our PHOENIX stellar model is interpolated between the spectral grid points presented in \cite{Husser2013} for the effective temperature $T_{\mathrm{eff}}$, surface gravity log $g$, and metallicity [Fe/H] values listed for ups~And~A in Table~\ref{systemproperties}. We rotationally broaden this model assuming a stellar rotation rate of 9.62 km/s \citep{Valenti2005} and limb darkening coefficient of 0.29 \citep{Claret2000}. For completeness, we instrumentally broaden the stellar model with the kernel determined in Section \ref{reduction}.

\subsection{Model Planetary Spectrum}
\label{planetmodel}
We compute a high-resolution (R=250,000) thermal emission spectrum of ups~And~b according to the SCARLET framework \citep{Benneke2015}. The thermal structure and equilibrium chemistry of the ups~And~b model spectrum are dependent upon the expected stellar flux at the location of the planet. The model assumes perfect heat redistribution (perhaps a flawed assumption, see \citealt{Crossfield2010} and Section \ref{atm}) and a solar elemental composition \citep{Asplund2009}. The temperature profiles are computed self-consistently for a 1 x solar, C/O=0.54 atmosphere by iteratively recalculating the radiative-convective equilibrium and atmospheric equilibrium chemistry. We assume an internal heat flux of $T_{int}$=75 K. Our default model in this paper is has an inverted temperature structure due to the short-wavelength absorption of TiO and VO. The SCARLET framework includes molecular opacities of $\mathrm{H_{2}O}$, $\mathrm{CH_{4}}$, $\mathrm{NH_{3}}$, HCN, $\mathrm{CO}$, $\mathrm{CO_{2}}$, and $\mathrm{TiO}$ (ExoMol database by \citealt{Tennyson2012}), molecular opacities of $\mathrm{O_{2}}$, $\mathrm{O_{3}}$, $\mathrm{OH}$, $\mathrm{C_{2}H_{2}}$, $\mathrm{C_{2}H_{4}}$, $\mathrm{C_{2}H_{6}}$, $\mathrm{H_{2}O_{2}}$, and $\mathrm{HO_{2}}$ (HITRAN database by \citealt{Rothman2009}), absorptions by alkali metals (VALD database by \citealt{Piskunov1995}), $\mathrm{H_{2}}$-broadening \citep{Burrows2003}, and collision-induced broadening from $\mathrm{H_{2}}/\mathrm{H_{2}}$ and $\mathrm{H_{2}/He}$ collisions \citep{Borysow2002}.

Line positions and amplitudes are critical to obtaining the correct cross-correlation function. We use the line information from ExoMol for $\mathrm{H_{2}O}$ and $\mathrm{CH_{4}}$. The line lists were computed using ab-initio calculations based on quantum mechanics. The line center wavelengths of these calculations are accurate. Line amplitudes are harder to compute in ab-inbitio calculations, but we are using the best state-of-the-art line lists available, which is ExoMol for the temperature encountered in hot Jupiters. Model spectra are convolved with the instrumental profile from Section~\ref{reduction} before the cross-correlation analysis.

\subsection{Two-Dimensional Cross Correlation}
\label{corr}
We use the TODCOR algorithm \citep{Zucker1994} to cross-correlate each order of data for each epoch with the stellar and planet models, yielding a two-dimensional array of cross-correlation values for different stellar and planetary velocity shifts.

As in \cite{Piskorz2016}, at this step, we eliminate the $K_r$ and $K_l$ band orders ranging from 2.3 - 2.4 $\mu$m from the analysis since there is high correlation between the stellar and planetary models themselves at these wavelengths. This means we remove any signal from carbon monoxide, and the dominant molecule in the planetary model in the remaining wavelengths is water vapor.

\begin{figure}[t]
\centering
\noindent\includegraphics[width=21pc]{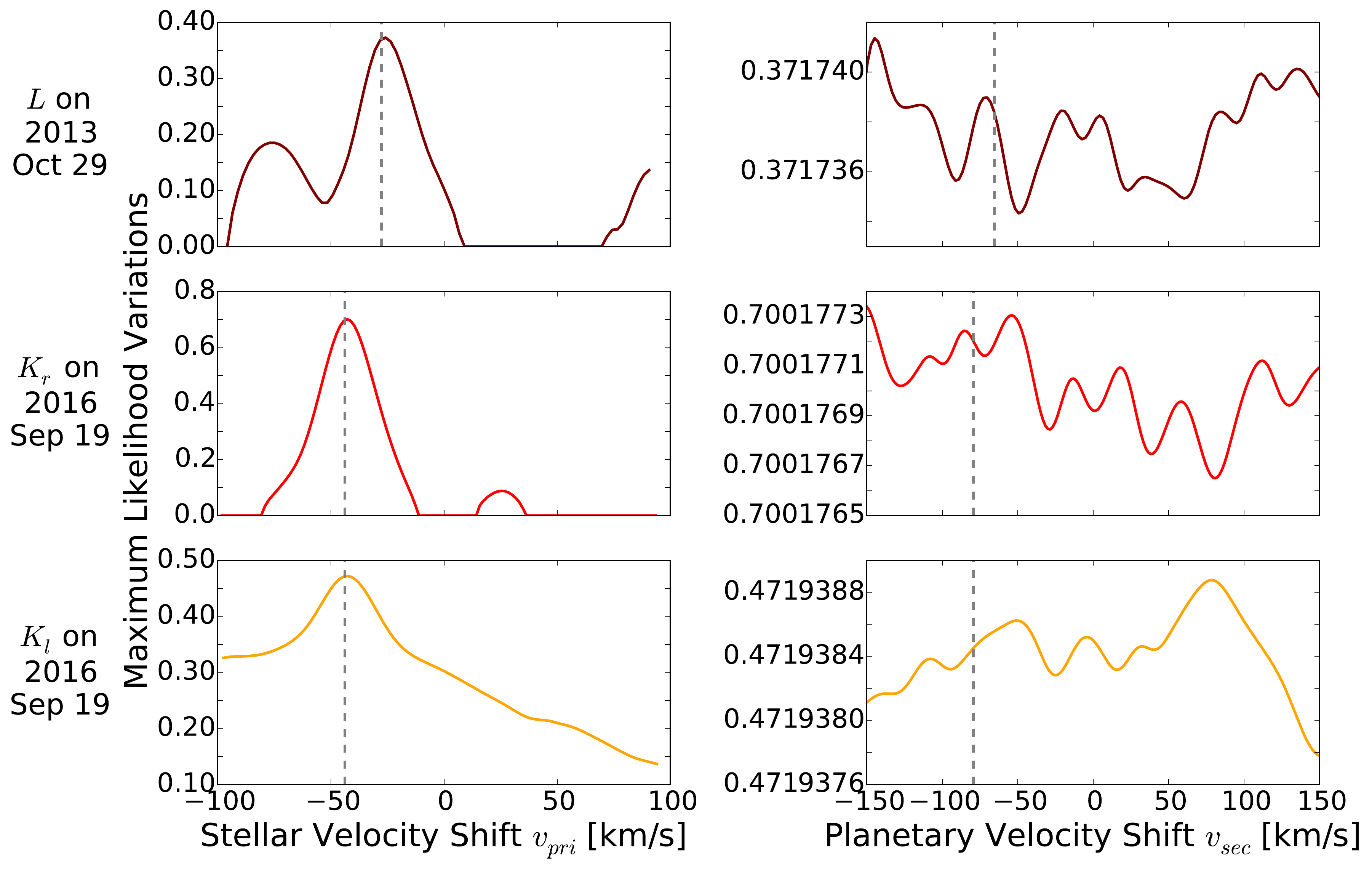}
\caption{Maximum likelihood functions for selected epochs of data in each band. Panels in the left column show the maximum likelihood function for the velocity shift of the star ups~And~in each band observed while panels in the right column shows the maximum likelihood function for the velocity shift of the planet ups~And~b. The grey vertical lines represent the expected values of $v_{pri}$ and $v_{sec}$ (based on the barycentric and systemic velocities and the line-of sight Keplerian velocity determined in Section \ref{orbitsoln}). Based on $\sigma_{f+\omega}$, the error on $v_{sec}$ is 0.4 km/s.}
\label{sxcorrL}
\end{figure}

Following \cite{Lockwood2014}, for each epoch of observations, we combine the correlation function for each order and produce nightly stellar and planetary maximum likelihood curves, a few of which are shown in Figure \ref{sxcorrL}. For every epoch, we are able to confirm the expected velocity of the star 
\begin{equation}
v_{pri} =v_{sys} -  v_{bary}
\end{equation}
(where $v_{sys}$ is the systemic velocity of ups~And~A and  $v_{bary}$ is the barycentric velocity of the Earth at the time of observation) as is the shown by the strong peaks in the panels on the left-hand side of Figure \ref{sxcorrL}. We suspect that the significant off-peak correlation signature in the primary velocity curve for the $K_l$ band data implies that we were too aggressive in our clipping and that we have scratched the noise limit of our data (see Section \ref{obsnotes}). 

The right column of Figure \ref{sxcorrL} shows the maximum likelihood curves for shifts in the planetary velocity. Two aspects are notable. First, the likelihood variations of the $K$ band data are an order of magnitude smaller than those of the $L$ data, indicative of the small signals present in the $K$ band data.  Second, there are many peaks and troughs in the planetary maximum likelihood curves. Therefore, determining the line-of-sight velocity of the planet is not straightforward. Only one peak in each maximum likelihood curve represents the real planetary velocity for a given epoch; the other peaks are chance correlations with the repeating structure in the planetary model. The multi-epoch data are critical in breaking this degeneracy.

\subsection{Planet Mass and Orbital Solution}
\label{orbitsoln}
\begin{figure*}[t]
\centering
\noindent
\includegraphics[width=42pc]{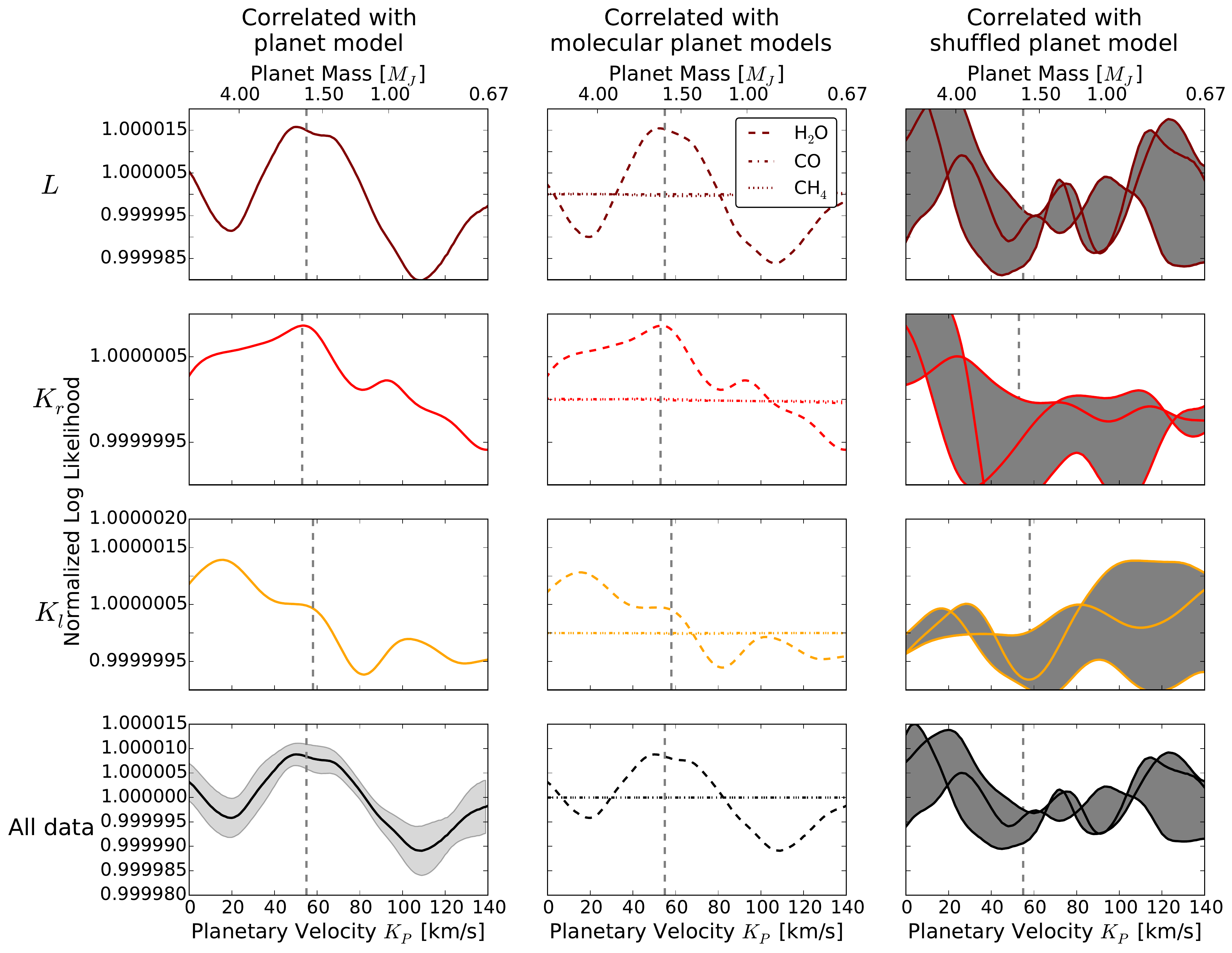}
\caption{Normalized log likelihood as a function of Keplerian orbital velocity $K_P$. Note that the vertical axes cannot be directly compared. Likelihood curves in the left column are the result of correlating NIRSPEC data with a SCARLET planet model for ups~And~b. The light shading on the maximum likelihood curve of all the data correlated with a planet model represent the jackknifed error bars. Likelihood curves in the center column are the result of correlation NIRSPEC data with SCARLET planet models containing single molecules. Likelihood curves in the right column are the result of correlation NIRSPEC data with multiple shuffled SCARLET planet models (which eliminates the planel signal in most cases); the dark shading is for the sake of clarity, only. The first row of likelihood curves considers only $L$ band data, the second only $K_r$ band data, the third only $K_l$ band data, and the fourth all the data.}
\label{kp}
\end{figure*}

We use the cross correlation functions for the planetary velocity shift $v_{sec}$ at each epoch to determine the most likely value of the line-of-sight Keplerian velocity $K_P$. For the sake of completeness, we use the equation for orbital velocity which considers eccentricity, even though the eccentricity of ups~And~b is very nearly zero. As a result of this near-zero eccentricity, the mean anomalies $M$ of our observations are essentially the same as the true anomalies $f$. The velocity $v_{sec}$ of the planet a function of its true anomaly $f$ is
\begin{equation}
\label{vf}
v_{sec}(f) = K_{p}(\cos(f+\omega)+e\cos\omega) + v_{pri} 
\end{equation}
where $K_P$ is the planet's orbital velocity, $\omega$ is the longitude of periastron measured from the ascending node, and $e$ is the eccentricity of the orbit. We test orbital velocities from -150 to 150 km/s in steps of 1 km/s and thus test a variety of planet masses and orbital inclinations. This results in a plot of maximum log likelihood versus the planet's orbital velocity (first column of Figure \ref{kp}).

Six $L$ band cross-correlation functions similar to that in the upper right panel of Figure \ref{sxcorrL} are combined to produce the likelihood curve in the upper left panel of Figure \ref{kp} when combined with equal weighting. The single peak in $K_P$ is at 55 $\pm$ 3 km/s. The error bars reported here are the three-sigma error on the mean value of a Gaussian curve fit to the maximum likelihood peak with equal weighting to the points on the maximum likelihood curve. The error bars are not the full-width at half-maximum of the fitted Gaussian. We more robustly calculate the weighting of the points on the maximum likelihood curve and the error bars and significance of the $K_P$ measurement based on the full eleven nights of data later in this section. Three $K_r$ band cross-correlation functions similar to that in the middle right panel of Figure \ref{sxcorrL} produce the likelihood curve in the second row of the first column of Figure \ref{kp} and shows a peak at $K_P$ = 53 $\pm$ 3 km/s. Finally, three $K_l$ band cross-correlation functions similar to that in the bottom right panel of Figure \ref{sxcorrL} produce the likelihood curve in the third row of the first column of Figure \ref{kp} and shows a peak at $K_P$ = 58 $\pm$ 3 km/s. 

The combination of all twelve nights of data is shown in the bottom left panel of Figure \ref{kp} and gives $K_P$ = 55 km/s. We use this value of $K_P$ to calculate the expected $v_{sec}$ for each epoch of observation and note this value as a vertical line on the curves in the right column of Figure \ref{sxcorrL}. For most cases (especially in $L$ and $K_r$ bands), the expected $v_{sec}$ corresponds to a local maximum in likelihood. We also use $K_P$ = 55 km/s to calculate the secondary velocities plotted in Figure \ref{rvplot}. 

Given the full suite of data, we calculate the error bars of each point of the maximum likelihood curve using jackknife sampling. We remove one night of data from the sample at a time and recalculate the maximum likelihood curve. The error on each point is proportional to the standard deviation of the twelve resulting maximum likelihood curves. These errors are shown in the bottom left panel of Figure \ref{kp}. These errors are an estimate only. For a Gaussian fit to the peak at 55 km/s, the reduced chi-squared value (chi-squared divided by the number of degrees of freedom) is 0.15, suggesting that the error bars are likely an overestimate. These large error bars are driven by a high variance in the jackknife samples. The Gaussian fit also gives error bars on the ultimate $K_P$ measurement: 55 $\pm$ 9 km/s.

To determine the significance of this detection, we use the jackknifed error bars to fit a Gaussian (above) and a straight line and the compare the likelihoods of the fits with the Bayes factor $B$. Here, the Gaussian fit corresponds to the presence of a planetary signal and the linear fit corresponds to the lack thereof. The Bayes factor $B$ is the ratio of the likelihood of two competing models \citep{Kass1995}. If 2ln$B$ is greater than 10, then the model is very strongly preferred.

For the Gaussian fit compared to the linear fit, the value of 2ln$B$ is 10.5, indicating that the signal at 55 km/s is stronger than a straight line at about 3.7$\sigma$. Therefore, the line-of-sight orbital velocity of ups~And~b is 55 $\pm$ 9 km/s. Using the indicative mass of ups~And~b and the law of conservation of momentum, we calculate that the true mass of ups~And~b is 1.7 $^{+0.33}_{-0.24}$ $M_J$, and the orbital inclination of ups~And~b is 24 $\pm$ 4$^{\circ}$.

\subsection{Measurements of ups~And~b's Atmosphere}
With SCARLET, we can calculate the contributions of individual molecules (H$_2$O, CO, and CH$_4$) to the total spectrum to understand the dominant opacity structures. We cross-correlate these molecular planet models with our $L$, $K_r$, and $K_l$ band data. Results of these single molecule cross-correlation calculations are shown in the middle column of Figure \ref{kp} for each band observed and indicate that the atmospheric opacity of ups~And~b is dominated by water vapor at the observed wavelengths. The likelihood curves for data correlated with CO- and CH$_{4}$-only planetary models show variations at least an order of magnitude smaller than the H$_2$O-only results. If carbon monoxide or methane are present at these wavelengths, they exists at levels below the detection limit of this data set. (See Section \ref{atm}.) Note that we were forced to remove the CO band at 2.2935 $\mu$m from our data set because of the presence of CO features in the stellar spectrum.

\section{Discussion}
\label{discussion}

\subsection{Tests of the Orbital Solution}
\label{tests}
Our initial test of the fidelity of the line-of-sight velocity detection at 55 km/s is to vary the spectroscopic contrast $\alpha_{spec}$. We test $\alpha_{spec}$ from 10$^{-7}$ to 10$^{-3}$ and find that the peak at 55 km/s is robust down to 10$^{-6.5}$.  $\alpha_{spec}$ is truly the ratio between the depths of the spectral lines, and so could be as low as zero for perfectly isothermal atmospheres.

Analagous to \cite{Piskorz2016}, we produce a ``shuffled" planetary model by randomly rearranging chunks of the planetary model. Cross-correlating our data with a shuffled model should show no peak near 55 km/s if the planet truly exists with a line-of-sight orbital velocity of 55 km/s. For each band of data, we run this test three times and the results are shown in the right-hand column of Figure \ref{kp}. The $L$, $K_r$, and $K_l$ band detections show minima near 55 km/s, showing that the planet signal is successfully eliminated. 

We use our inclination measurement of 24 $\pm$ 4$^{\circ}$ to compare our detection to the results presented in other works. The spectroscopic technique presented here would be unable to detect the motion of ups~And~b if $i_b<$ 4.9$^{\circ}$ due to the size of a resolution element on NIRSPEC. Our inclination measurement is largely in agreement with previous works. Spitzer brightness measurements indicated $i_b>$ 28$^{\circ}$ \citep{Crossfield2010}. Newtonian orbital simulations considering the orbital elements of ups~And~c and d suggested that orbits having $i_b<$ 60$^{\circ}$ can be stable \citep{McArthur2010}.  Analagous post-Newtonian orbital simulations prescribed a ``region of stability" for $i_b<$ 40$^{\circ}$. Our measurement of $i_b$ = 24 $\pm$ 4$^{\circ}$ lies within the error bars of these ranges.

\subsection{System Stability}
Many previous works have characterized the ups~And~A system as on the edge of instability. Here we evaluate our calculation of the inclination of ups~And~b by running numerical simulations of the system with the Mercury software \citep{Chambers1999}. Mercury is a hybrid-simplectic--Burlisch Stoer algorithm \citep{Chambers1999}. We include the central star ups~And~A and the three planets ups~And~b, c, and d, set the time step to one-twentieth of the orbital period of ups And b, and consider general relativity.  

Our method of calculating $K_P$ and $i_b$ provides no insight into the longitude of ascending node of ups~And~b, $\Omega_b$. As a result, we investigate values of $i_b$ between 22$^{\circ}$ and 27$^{\circ}$ in steps of 1$^{\circ}$ and values of $\Omega_b$ between 0$^{\circ}$ and 360$^{\circ}$ in steps of 10$^{\circ}$. We adjust $M_b$ as is necessary given the value of $i_b$. All other orbital elements are taken from \cite{McArthur2010}. Specifically, the orbital elements used for our simulations are listed in Table \ref{systemproperties}.

\begin{figure}[t]
\centering
\noindent\includegraphics[width=22pc]{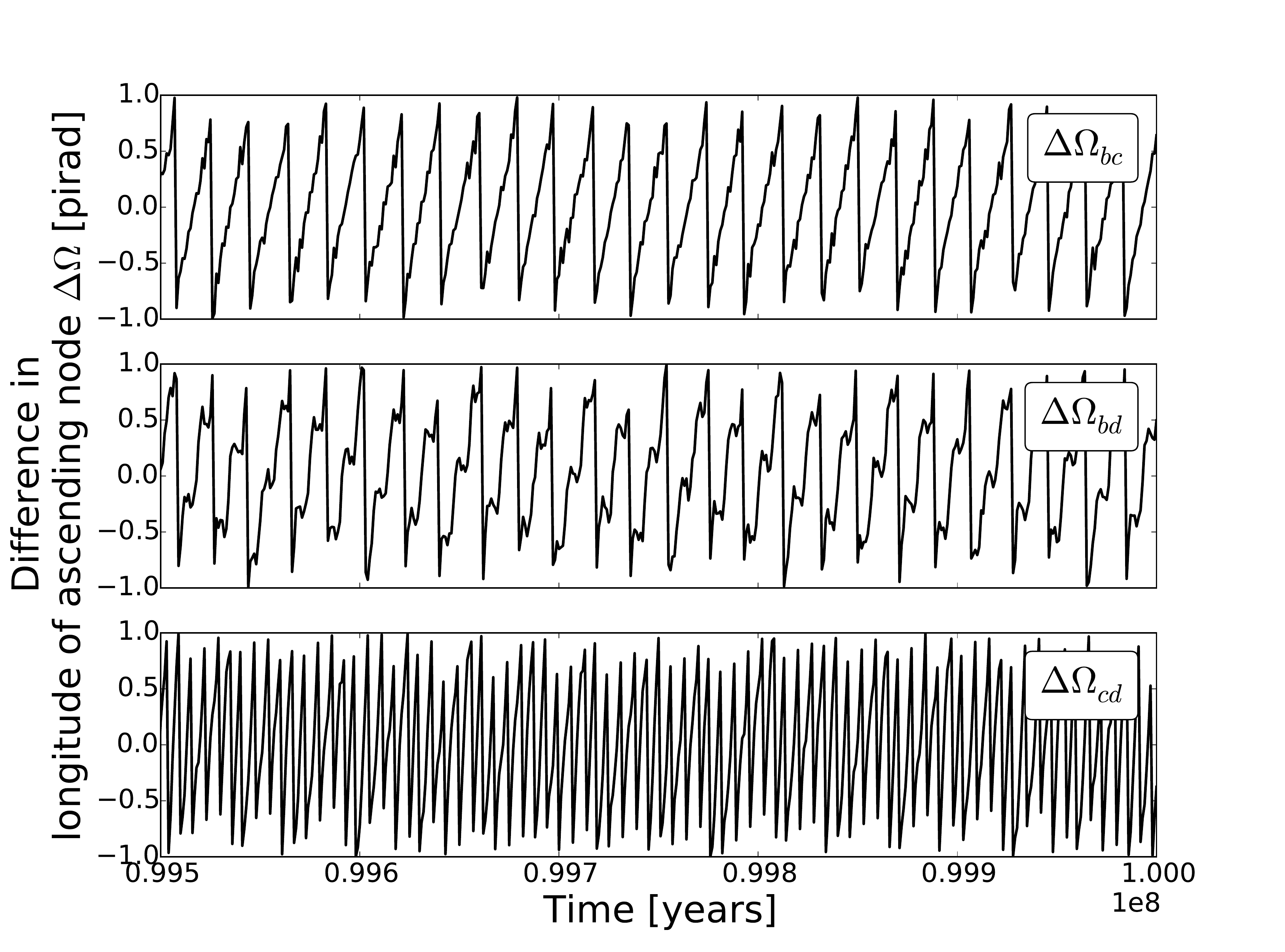}
\caption{Plot of the difference in longitude of ascending node $\Delta\Omega$ versus time for the last 500,000 years of the 100 Myr Mercury simulation for each pair of planets in the ups~And~system. This simulation was initialized with $i_b$ = 24$^{\circ}$ and $\Omega_b$ = 0$^{\circ}$.}
\label{deltaomega}
\end{figure}

Of our 216 simulations, 122 were stable for more than 100,000 years. These simulations have $\Omega_b <$ 100$^{\circ}$ or $\Omega_b >$ 260$^{\circ}$.  Of these systems, 53 were stable for more then 1 Myr, having $\Omega_b <$ 40 $^{\circ}$ and $\Omega_b >$ 320$^{\circ}$. We extract the 24 simulations having 23$^{\circ}< i < 25^{\circ}$ and run them for 100 Myr.  All but two are stable. It seems that for the successful simulations the orbital planes of planets b and d remain roughly aligned. For example, if $i_b$ = 24$^{\circ}$ and $\Omega_b$ = 0$^{\circ}$, then the mutual inclination of  ups~And~b and c is about 29$^{\circ}$ and the mutual inclination of  ups~And~b and d is about 2$^{\circ}$. Recall, the mutual inclination of ups~And~c and d is 29$^{\circ}$. Successful simulations tend to have mutual inclinations clustered about these values. In these simulations, the apsides of ups~And~c and d oscillate as in \cite{Chiang2002}, \cite{Barnes2011}, and other works, and the orbital evolution of ups~And~b is secular (Figure \ref{deltaomega}). We stress that these simulations are stable not necessarily because of the value of ups~And~b's inclination, but because of the direction ups~And~b's inclination vector points over time. Our Mercury simulations provide evidence that stable ups~And~A systems do indeed exist for the inclination we have measured, and provide insight into the three-dimensional geometry of ups~And~b's orbit.


\subsection{The Atmosphere of ups~And~b}
\label{atm}
In our planetary model, the $L$ band opacity is dominated by water vapor. Therefore, our $L$ band detection of ups~And~b's thermal emission spectrum suggests that radiative transfer in the planet's atmosphere is dominated by water vapor at these wavelengths. In fact, the source of the correlation signal for all wavelengths investigated is water vapor (see the middle column of Figure \ref{kp}). Based on the analysis of $\alpha_{spec}$ presented in Section \ref{tests}, the detection of H$_2$O suggests that its spectroscopic contrast $\alpha_{spec} >$ 10$^{-6.5}$.

We perform a comparison of the cross-correlation results given inverted and non-inverted model spectra. The main differences in the final maximum likelihood curves stem from the different line strengths at a given wavelength for each model. In other words, the differences stem from the optical depths as a function of wavelength. Therefore, the only conclusion we can draw at this time is the atmosphere of ups~And~b is dominated by water at the probed wavelengths.

Though the $K$ band is typically dominated by CO absorption, the usable $K$ band wavelengths in our dataset do not include strong CO absorption. The non-detections of CO and CH$_4$ suggest that their spectroscopic contrasts are $\alpha_{spec} <$ 10$^{-6.5}$ at these wavelengths.

Our models do not account for cloud cover, atmospheric recirculation, or the differences between dayside and nightside spectra. \cite{Crossfield2010} reported a flux maximum in the Spitzer phase curve of ups~And~b at 80$^{\circ}$ before opposition, or at mean anomaly $M$ = 0.4, in our formulation. $M$ = 0.4 is almost directly between the phases of 2016 November 12 and 2016 August 21 observations as diagrammed in Figure \ref{schematic}. Fortuitously, this indicates that even if the planet's flux maximum is shifted from what would traditionally be expected, our measurements are still able to capture dayside emission.

\subsection{Observation Notes}
\label{obsnotes}
\begin{figure}[t]
\centering
\noindent\includegraphics[width=22pc]{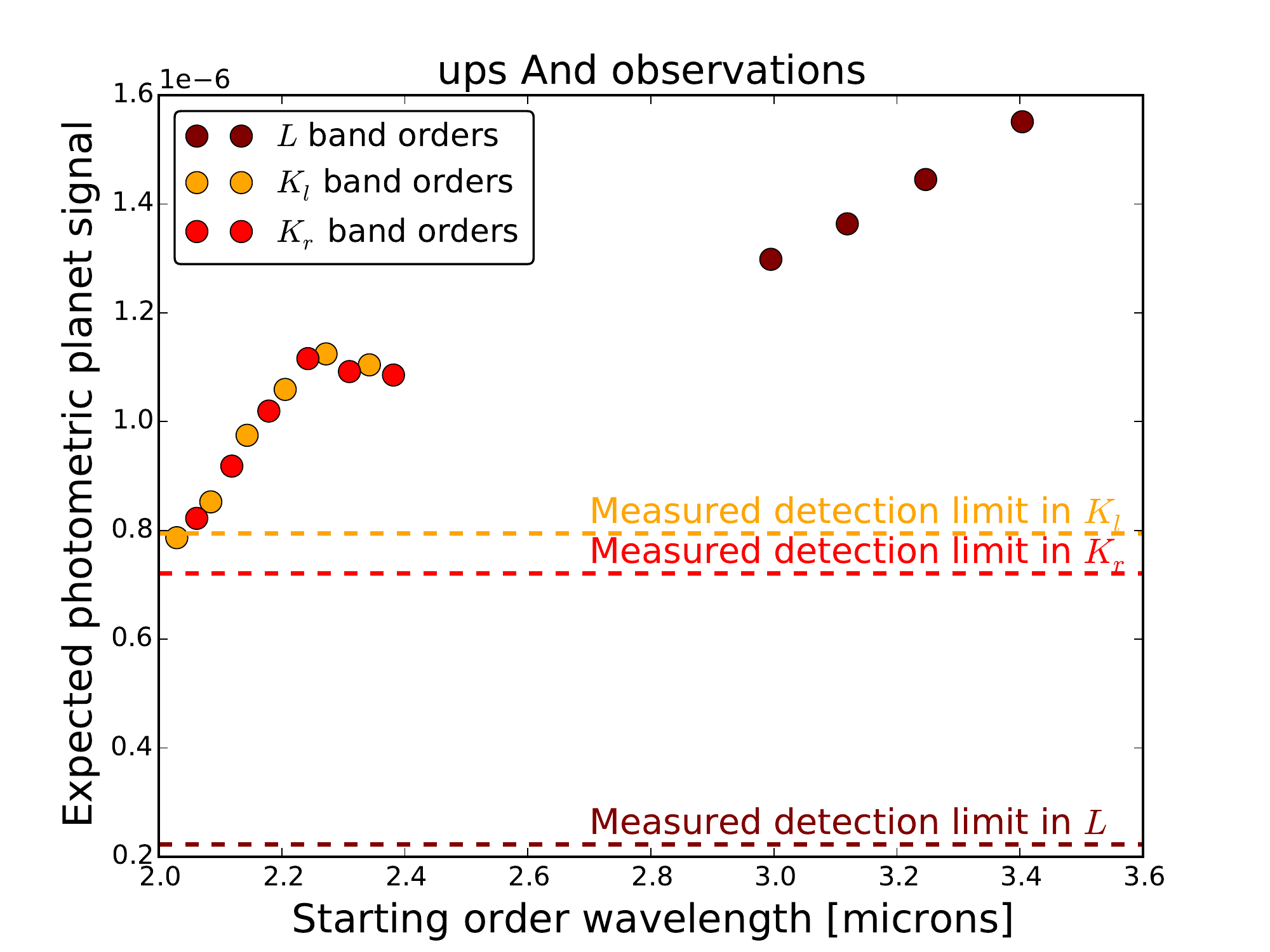}
\caption{Expected planet-star contrast as a function of starting order wavelength compared to achieved photometric contrast. Points represent the expected planet photometric signal calculated from a PHOENIX stellar model and a SCARLET planet model for each observed order of data (6 orders in $K_l$, 6 orders in $K_r$, and 4 orders in $L$). Dotted lines represent the achievable contrast given by the aggregate shot noise for all epochs of data in each band. Note that in our analysis we do not use the first two and final orders of the $K_l$ and $K_r$ bands.}
\label{contrast}
\end{figure}

From our raw data sets, we calculate the shot noise per resolution element for each observation. (See Table \ref{observationtable}.) We compare the aggregate shot noise values to the expected photometric signal from the planet for each order observed, using the stellar and planet models described in Sections \ref{stellarmodel} and \ref{planetmodel}. As Figure \ref{contrast} suggests, we easily achieve the required S/N to detect the planet with six nights of $L$ band observations, but only marginally achieve that required with three nights of $K_l$ and $K_r$ observations. In fact, we achieve slightly better shot noise for $K_r$ than for $K_l$, a possible reason for the stronger detection of the planet signal here (Figures \ref{sxcorrL} and \ref{kp}).

In this suite of observations, the $K_l$ band data sets are equivalent to the $K$ band data presented for HD 88133 in \cite{Piskorz2016}. With four nights of $K$ band data, \cite{Piskorz2016} were able to detect the signal from HD 88133 b, though not as clearly as in the six nights of $L$ band data. This points to the general trend that, with NIRSPEC at Keck, $L$ band observations may be more amenable to direct detection of exoplanet atmospheres than those in the $K$ band. For hot Jupiters, the increase in the thermal background from $K$ to $L$ band is more than compensated for by the significant increase in planet flux relative to the star. In other words, though the increment of detection limit achieved per unit integration time is higher in the $K$ band than in the $L$ band, the star-planet contrast near 2 $\mu$m may be too small for a bona fide planet detection with our data. For this cross-correlation method, the superiority of $L$ band observations over $K$ band observations is a demonstration of the theoretical results presented in \cite{deKok2014}.

\section{Conclusion}
We detect the thermal emission spectrum of ups~And~b with ground-based high-resolution spectroscopy. For the hot Jupiter ups~And~b, we find a Keplerian velocity of 55 $\pm$ 9 km/s, a true mass of 1.7 $^{+0.33}_{-0.24}$ $M_J$, and an orbital inclination of 24 $\pm$ 3. We show that the ups~And~A system is stable for at least 100 Myr given the reported ups~And~b orbital elements. Using the many planet lines available in the $L$ and $K$ bands, we determine that the planet's opacity structure is dominated by water vapor. For the set of observations presented here, the signal is noticeably stronger in the $L$ band than in $K$, suggesting that $L$ band observations may be best suited for these analyses moving forward. Further thermal IR measurements can be used to dig deeper into the structure and compositions of hot Jupiter atmospheres and eventually atmospheres of planets at larger semi-major axes.

\acknowledgments{The authors thank an anonymous reviewer for useful comments and suggestions on this paper. The authors thank Konstantin Batygin for guidance and insight into the stability of this planetary system. The authors wish to recognize and acknowledge the very significant cultural role and reverence that the summit of Mauna Kea has always had within the indigenous Hawaiian community.  We are most fortunate to have the opportunity to conduct observations from this mountain. The data presented herein were obtained at the W.M. Keck Observatory, which is operated as a scientific partnership among the California Institute of Technology, the University of California and the National Aeronautics and Space Administration. The Observatory was made possible by the generous financial support of the W.M. Keck Foundation. This work was partially supported by funding from the NSF Astronomy \& Astrophysics and NASA Exoplanets Research Programs (grants AST-1109857 and NNX16AI14G, G.A. Blake P.I.). Basic research in infrared astrophysics at the Naval Research Laboratory is supported by 6.1 base funding.}

\end{document}